\documentstyle{article}

\newcommand{\mathbf}{\bf}

\begin{document}

\begin{center}
{\huge\bf On The relation between Superconductivity and Quantum  
Hall Effect}
\end{center}

\vspace{1cm}
\begin{center}
{\large\bf
F.GHABOUSSI}\\
\end{center}

\begin{center}
\begin{minipage}{8cm}
Department of Physics, University of Konstanz\\
P.O. Box 5560, D 78434 Konstanz, Germany\\
E-mail: ghabousi@kaluza.physik.uni-konstanz.de
\end{minipage}
\end{center}

\vspace{1cm}

\begin{center}
{\large{\bf Abstract}}

We introduce a model of superconductivity and discuss its relation  
to the quantum Hall-effect. This kind of relation is supported by  
the well known SQUID results. The concept of pure gauge potential as  
it is involved in various theoretical models concerning solid state  
effects in magnetic fields is also discussed.
\end{center}

\begin{center}
\begin{minipage}{12cm}

\end{minipage}
\end{center}

\newpage
The main properties of superconductors are according to \cite{all  
1} that the magnetic field does not penetrate into the  
superconcuctor, i. e. $B_{in} = 0$ and that no macroscopic volume  
current can flow in a superconductor. Thus any electric current  
which flows in a superconductor must be a surface current.
These properties refer to a $2+1$-D theory of superconductivity,  
where the electromagnetic potentials are pure gauge potentials  
\cite{gauge}, i. e. with vanishing field strength $F$, $B\in F$  
which should be described by a Chern-Simons-action functional.
However, one must keep in mind that $B_{in} = 0$ or any vanishing  
of conjugate varibles in quantum mechanics have a limit according to  
the uncertainty relations \cite{ unb}.
Morerover, considering the short range photons in Meissner-effect  
\cite{all 1} and the fact that in the Maxwell-Chern-Simons-action in  
$2+1$-D the photons become massive and short ranged \cite{vier}  
(see also Ref. [1f]), we are led  again to the $2+1$-D theory of  
superconductivity. Furthermore, note that the most importent hint  
about the pure gauge character of the electromagnetic potential in  
superconductivity comes from the coherent phase of the wave function  
of the whole system of the Cooper-pairs in the BCS-theory \cite{all  
1}. Hence, its phase is given by the line integral of the  
electromagnetic potential which means that this one is a pure gauge  
potential which results in the flux quantization. This shows  
definitely that the main concept of the BCS-theory is the pure gauge  
potential describing the main object of the theory, i. e. the wave  
function of Cooper-pairs \cite{noph}. Moreover, the possibility of  
the Cooper-pairs is a quantum mechanical result in $2+1$ dimensions,  
i. e. in a $2$-D potential [1d], \cite{neu}. In view of the  
accepted fundamentality of the BCS-theory this is enough hint about  
the fundamentality of the mentioned $2+1$-D point of view which  
contains also the pure gauge conception.

\medskip
The usual theory of superconductivity is based on modifications of  
the Maxwell's electrodynamics according to the phenomenological  
London's equations, or according to the Ginzburg-Landau-model which  
are belived to have their microscopic explanation in the BCS-theory.  
Nevertheless, also the fundamental concept of the  
Ginzburg-Landau-theory is the $2$-dimensional pure gauge potential  
from which one obtains not only the Ginzburg-Landau-equations but  
also the Landau's coherence-length \cite {all 1} \cite{gauge}. In  
the same manner also the main concept in London's equations is that  
of the pure gauge potential which results in the flux quantization  
\cite{gauge}. Furthermore, also in case of the Josephson effect  
which is considered beyond of the flux quantization as a  
manifestation of the BCS-theory, a new general approach contains the  
$2$-dimensional pure gauge potential \cite{rus}. Thus, also the  
various sub-models and sub-effects of the superconsuctivity and weak  
superconductivity, i. e. superconductivity in lower or higher  
magnetic fields all demonstrate variations of the pure gauge  
potential picture. Furthermore, beyond the mentioned methodes most  
of other perturbative technics in solid state physics like the  
methodes of Green's function or of pseudo-potentials contain also  
the concept of pure gauge potential {\it in multiply connected  
regions}. Close to this concept it is related the concept of multy  
valued functions and the relation is based on the invariant  
relations between the topological properties of manifolds and the  
invariant properties of functions defined on the manifolds, e. g.  
via Morse-theory, harmonic functions, etc.

On the other hand, consider the fundamental differential geometric  
discrepancy between the Maxwell's equations $d^{\dag} F = j$ and the  
London's equations $ F = dj $. Recall also the discrepancy with the  
Ginzburg-Landau's equations for higher magnetic fiels which are  
given by $d^{\dag} F = J$.
Furthermore, considering the Ohm's equations $F = *\rho J$ with  
$\rho$ as the resistivity matrix \cite{all 2}, we have again a  
discrepancy with the Maxwell's equations. Here $F$ and $J$ are the  
field strength and current density two-forms respectively, whereas  
$j$ is the current density one-form. Thus, in all effects concerning  
the conductivity we need additional "phenomenological" relations  
for currents and conductivity usually called the "material  
equations" beyond the elaborated Maxwell's equations. Nevertheless,  
one must keep in mind that $J$ and $j$ are refering to two different  
current densities, namely to the electromagnetic and to the  
electric ones respectively (see below). The difference is just the  
electromagnetic potential which appears in these models as a pure  
gauge potential \cite{gauge}.

These are discrepancies arising from the hirarchy of the $3+1$-D  
consideration of conductivity which refer to a possible solution of  
discrepancies within a $2+1$-D model. Of course, related with these  
questions are also the hirarchy of the band-theory and phonons in  
the solid state physics. According to some new field theoretical  
approaches to quantum Hall-effect (QHE) and superconductivity it  
seems that the general foundations of these effects does not reflect  
any band-structur properties \cite{QCN Bk}\cite{LafSC} \cite{noph}.  
Furthermore, there are the so called non-phonon mechanisms models  
which also explain the high-$T_c$ superconductivity according to the  
pure electronic behaviour of superconductors \cite{noph}.  
Therefore, one can expect that it should be possible to relate not  
only the high $T_c$ superconductivity but the whole  
superconductivity with QHE for which there are already non-phononic  
models \cite{QCN Bk}\cite{mein}. One advantage of this approach will  
be the possibility to understand the notion of the critical  
magnetic induction $H_c$. Hence, according to a possible unified  
$2+1$-D approach to QHE and superconductivity, if the exterior  
magnetic field increases the value of $B_c$ then the  
superconductivity of the particle system proceeds into the QHE which  
should be observable if the system is prepared in the proper way  
(see below).  Hence, there were already hints in this direction  
according to which the ground state of a system of non-interacting  
particles with fractional statistics can be considered as a new kind  
of high-$T_C$ supercunductor \cite{LafSC}.Thus, there are enough  
hints about the neccesity of a $2+1$-D electrodynamical approach to  
the superconductivity \cite{noph}.

Recently, we showed that also a microscopic theory of the integer  
quantum Hall-effect (IQHE) is possible within a  
Chern-Simons-Schroedinger theory without any relation to the  
band-theory \cite{mein}. Furthermore, there are already models of  
fractional Hall-effects (FQHE) of the same Chern-Simons-type  which  
are also independent of the band-theory \cite{modF}. Thus, there  
should be beyond the usual solid state theoretical approaches to the  
conductivity a "parallel" $2+1$-D gauge field theoretical approach  
which should incorporate also the Chern-Simons electrodynamical  
descriptions of the conducting systems. In view of the fact that the  
usual, i. e. the semi-classical conductivity, the quantum  
Hall-effects (IQHE and FQHE) and the superconductivity (low and  
high-$T_c$) are all various forms of the electromagnetic  
cunductivity, it seems to be natural to look on a general theory of  
cunductivity connecting all of its features.

\medskip
Here we introduce a brief theoretical model which incorporates the  
main electromagnetic aspects of the superconductivity and of the  
QHE, where we consider both systems as $2+1$-dimensional systems  
\cite{mein} \cite{modF} \cite{kurz}.

To begin recall that the varification of the BCS-theory of  
superconductivity  follows from the flux quantization and the flux  
quantization is a result of pure gauge field character of the  
electromagnetic potential \cite{gauge}. More precizely, the flux  
quantization is the "quantum mechanical" or global, i. e. invariant  
expresion of the pure gauge character of the electromagnetic  
potential involved in the electromagnetic current (see below)  
\cite{gauge}. This is in accordance with the above mentiond fact  
that for a superconducter $B_{in} = 0$. It is in view of the global  
character of the wave function in quantum mechanics that locally  
vanishing effects which are therefore non-observable in the  
classical level for example by Lorentz-force equations, become  
observable in quantum mechanics.
Furthermore, the local pure gauge property of an electromagnetic  
potential is given by the local expresion of the vanishing of  
electromagnetic current density, i. e. $J_{m}^{(em)} =  
{\displaystyle\frac{-ie}{2M_e}(\psi^*(\partial_m \psi) - (\partial_m  
\psi^*) \psi) - \frac{e^2}{M_e}\psi^*A_m\psi} = 0$ for all $\psi$  
\cite{all 1}, where $M_e$ is the mass of electron and we set $\hbar  
= 1$ and $m,n = 1,2$ in $2+1$ dimensions.
Moreover, one can obtain also the London's equations from the same  
vanishing of electromagnetic current density.

Now let us observe that, the proper current density for the  
superconductivity described by the London's equations is the {\it  
electric} current density which is valid also in presence of low  
magnetic fields, i. e. for $\omega_c \tau \ll 1$. It is given by  
$j_{m}^{(e)} = {\displaystyle\frac{-ie}{2M_e}(\psi^*(\partial_m  
\psi) - (\partial_m \psi^*) \psi)}$. On the other hand in case of  
QHE, i. e. in presence of higher magnetic fields or in the limit  
$\omega_c \tau \gg 1$, the {\it electromagnetic} current density is  
given by $J_{n}^{(em)} = j_{n}^{(e)} -  
{\displaystyle\frac{e^2}{M_e}\psi^* A_n\psi}$. Thus, if we set, as  
usual \cite{all 1} the $J_{n}^{(em)} = 0$

\begin{equation}
j_{n}^{(e)} - \frac{e^2}{M_e}\psi^* A_n\psi = 0
\end{equation}
\label{pure}

These could be considered as the local solutions of London's  
equations or equivalently as the London's equations themselves  
\cite{all 1}. Thus, the time and exterior spatial derivative of  
equations (1) result in the original version of London's equations  
mentioned above. On the other hand, if we consider the $V_{\alpha} =  
J_{\alpha}^{(em)} = j_{\alpha}^{(e)} -  
{\displaystyle\frac{e^2}{M_e}\psi^*A_{\alpha}\psi}$ with  
${\{\alpha,\beta,\gamma}\} = {\{1,2,3}\}$ as a $2+1$-D vector  
potential which obey the Lorentz-condition in view of the continuity  
equation for  $J_{\alpha}^{(em)}$, then its dynamics following its  
pure gauge character according (1) should be given by a  
Chern-Simons-action:

\begin{equation}
\int \epsilon^{\alpha\beta\gamma}J_{\alpha}^{(em)}\partial_{\beta}  
J_{\gamma}^{(em)}\;\;\;,
\end{equation}
\label{wgl}

Hence, the constraint equations of this action are given by the  
relation (1).

Thus, the equations of motion and the constraint equations of the  
action (2) are the London's equations  in $2+1$ dimensions $  
\epsilon^{\alpha\beta\gamma} F_{\beta\gamma} (V) =  
\epsilon^{\alpha\beta\gamma}\partial_{\beta} J_{\gamma}^{(em)} = 0$.  
Therefore, in view of the gauge invariance of the electromagnetic  
field strength, London's equations are the local gauge invariant  
expresion of the pure gauge character of the electromagnetic  
potential.

On the other hand, it is known also that QHE \cite{all 2} is also  
related with the pure gauge character of the electromagnetic  
potential expressed by the Chern-Simons-action \cite{mein}.
We showed \cite{mein} also that the characteristic currents of  
IQHE, namely its edge currents result also from the pure gauge  
character of the electromagnetic potential in IQHE according to the  
constraints of the theory under the typical integer quantum Hall  
conditions \cite{kk}. Therefore, it seems plausible that the  
superconductivity and QHE become related with each other in view of  
the fact that they manifest various properties of the  
electromagnetic pure gauge  potentials.

Recall further that in the language of differential geometry the  
Londons equations \cite {all 1} in $2+1$ dimensions are given by $dj  
= (\lambda)^{-1} dA$, where $j$, $A$ and $\lambda$ are current  
density one-form, the gauge potential one-form and the London's  
penetrating depth $\lambda = {\displaystyle\frac{M_e}{n e^2}}$  
respectively, if we set $\mu_0 = 1$.
On the other hand the Ohm's equations \cite{all 2} for Hall-effect  
are given also in $2+1$-dimensions by $J^{(em)} = \sigma dA$, where  
$\sigma$ is the conductivity matrix. Furthermore, in $2+1$  
dimensions we can introduce a geometrical current density $J =  
*\lambda dj^{(e)}$. Hence, $J$ obeys also the continuity relation  
$*\partial* J = 0$. Thus, we could have a formal relation $J = *dA$  
which is similar to the Ohm's equations of QHE. Recall also that  
$J^{(em)}$ and $j^e$ both have the dimension $L^{-2}$ in accordance  
with the particle density on a surface.

It is obvious from such a constelation that the geometrically  
introduced current density $J$ can be identified with the current  
density of the Ohm's equations $J^{(em)}$ only if the conductivity  
marix $\sigma$ becomes equal to the $SO(2)$ matrix for the  
superconducting case which characterizes the rotation \cite{rot}. We  
will show that this condition which is equivalent to $\sigma_H = 1$  
is ideed the case corresponding with the mentioned ground state of  
the quantum Hall-system \cite{LafSC} which includes also the  
empirical fact of vanishing of longitudinal conductivity in QHE  
\cite{kk}.

\medskip
The Chern-Simons-action functional of our model for  
superconductivity which is a slight generalization of the action (2)  
and from which we can obtain the London's equations as the  
equations of motion is the following one defined on a $2+1$-D  
manifold $M = \Sigma\times\mathbf R$ with boundary.

\begin{equation}
\int \epsilon_{\alpha\beta\gamma} C_{\alpha} \partial_{\beta}  
C_{\gamma}\;\;\;,
\end{equation}
\label{cs}

with $C_{\alpha} = \lambda j_{\alpha}^{(e)} - \sigma_H A_{\alpha}$,  
where the $\sigma_H$ is the locally constant dimensionless  
parameter called the Hall-conductivity. The $j_{\alpha}^{(e)}$ is  
the usual electric current density of a {\ it non-interacting}  
system of electrons.

\medskip
If we set the value of the Hall-cunductivity $\sigma_H = 1$ to  
identify the ground state, then the equations of motion for  
$A_{\alpha}$ or $C_{\alpha}$:

\begin{equation}
\epsilon^{\alpha\beta\gamma} \partial_{\beta} C_{\gamma} = 0 \;\;\;\;,
\end{equation}
\label{sol}

are devided into the equations of motion for $C_m$

\begin{equation}
\frac{dj_m^{(e)}}{dt} = \lambda^{-1} \frac{dA_m}{dt} \;\;\;,
\end{equation}
\label{eqmot}

and the constraint equation given by

\begin{equation}
\epsilon_{mn} \partial_m j_n^{(e)} = \lambda^{-1}B \;\;\;, \;\;\;B  
= - \epsilon_{mn}\partial_m A_n \;\;\;,
\end{equation}
\label{const}

with $\epsilon_{mn} = - \epsilon_{nm} =1$ and ${m,n} = {1,2}$.

These two groups of equations are the London's equations in  
$2+1$-dimensions.

\medskip
To see the relation of $\sigma_H = 1$ condition with  
superconductivity let us recall that according to the definition  
$\sigma_H := \displaystyle{\frac{\sigma_0 \omega_c \tau}{ 1 +  
(\omega_c \tau)^2}}$ \cite{all 2} if the $\sigma_H = 1$ in the  
quantum Hall-limit of IQHE where $\omega_c \tau \gg 1$, then the  
electric conductivity $\sigma_0$ becomes equal to $\omega_c \tau$ or  
$\sigma_0 \gg 1$. Followingly, the electric resistivity becomes  
very small indicating the superconductivity. It is intresting to  
mention that also in the classical Hall-limit, i. e. under the  
classical Hall-conditions or
$\omega_c \tau \ll 1$, the $\sigma_H = 1$ condition is related also  
with large electric conductivity $\sigma_0 = (\omega_c \tau)^{-1}$  
or with the superconductivity. This posibility is related with the  
interplay between the strength of the electric and magnetic fields  
and the mean free time or mobility or the temperature which is  
correlated with the relation between the high-$T_c$  
superconductivity and the FQHE \cite{LafSC} where the mobility plays  
empirically an importent role. Furthermore, if $\sigma_H$ in (3)  
becomes equal to any positiv fractional number indicating the FQHE  
situation, $\sigma_0$ becomes according to the definition of  
$\sigma_H$ again proportional to $(\omega_c \tau)^{-1}$ which  
represents in view of $\omega_c \tau\gg 1$ again the  
superconductivity \cite{N}. Thus, for decreasing magnetic fields  
also the FQHE situation can result in the superconducting case if  
the preparation of the system allows \cite{LafSC}. Therefore : 1.)  
The usual superconductivity described by the standard London's  
equations should be related with the $\sigma_H = 1$ case of IQHE  
with non-interacting prticles. 2.) The high-$T_c$ superconductivity  
should be related with the ground state of the non-interacting  
particles in FQHE {\it after} decreasing the magnetic fields  
\cite{N}.

\medskip
It is importent to mention that the empirical information from  
superconducting quantum interfrence devices (SQUID) support this  
point of view. According to Ginzburg \cite{Ginz} the diagram of the  
magnetic flux $\Phi$ through the superconducting ring (with a weak  
link) with respect to the variation of the external magnetic flux  
has quasi plateaus on the integer magnetic flux $\Phi = \phi_0,  
2\phi_0, 3\phi_0, ...$. This diagram is similar to the digram of  
Hall conductivity with respect to the filling factor in IQHE [9].  
Roughly speaking, one could understand such a similarity, in view of  
the multiply connectedness of region in both cases and the  
essential role played by the pure gauge potential therein, if one  
relates the quantization of the Hall-conductivity $\sigma_H = \nu  
{\displaystyle {\frac{e^2}{h}}}$ where $\nu$ is the filling factor  
with the flux quantization according to $B_{(SC)}\cdot S = {\mathbf  
Z}{\displaystyle {\frac{h}{e}}}$ in superconductivity (SC), where  
$S$ is the area surrounded by the ring and ${\mathbf Z}$ is an  
integer. The relation becomes obvious if one recalls that under QHE  
conditions
$\omega_c \tau\gg 1$ one has
$\sigma_H = {\displaystyle \frac{ne}{B_{(QHE)}}} = {\displaystyle  
{\frac{Ne}{B_{(QHE)}\cdot S^{\prime}}}}$, where $B_{QHE}\gg B_{SC}$  
and $S^{\prime}\ll S$ is the area of the ring itself where the  
charge carriers are. To be precise it has to be mentioned that  
$S^{\prime}$ is the area of a ring with a width of $l_B$, so that if  
$S = \pi R^2$ is the empty area, then $S^{\prime} = 2 \pi R\cdot  
l_B$ is the ring-area or the edge-area where the edge current should  
move around.
Now, if $B_{(QHE)}\cdot S^{\prime} = B_{(SC)}\cdot S$, then the two  
quantization in two different levels of external magnetic field  
strength are correlated by $\nu = {\displaystyle {\frac{N}{\mathbf  
Z}}}$. In this manner, where integer and fractional $\nu$ implies  
IQHE and FQHE respectively, it is possible that QHE contains  
superconducting effect and vice versa. Moreover, one rediscover the  
well known empirical fact about the proportion
$\displaystyle{\frac{B_{(QHE)}}{B_{(SC)}}} =
\displaystyle{\frac{S}{S^{\prime}}} = \displaystyle{\frac{R}{2  
l_B}}$ in the mentioned Corbino-type samples between the height of  
magnetic field in QHE and superconductivity \cite{wert}. This  
relation can be considered also as an explanation for the  
quantization of Hall-conductivity $\sigma_H$ according to the flux  
quantization.

Such a relation between the QHE and Superconductivity should be  
realized by the Q-1-D superconducting systems \cite{bedno}. These  
systems demonstrate under large magnetic fields along the low  
conductivity axis an IQHE-behaviour \cite{bedno}.

Theoretically, this behaviour becomes clear in view of the fact  
that the quantization of the $2+1$-D field theories result in the  
$1+1$-D quantum theories of "chiral" currents \cite{verlinde}.
We showed, that the quantization of the $2+1$-D IQHE model results  
in the $1$-dimensional edge currents \cite{mein} which should be  
described dynamically by $1+1$-D quantum theories \cite{verlinde}.  
Hence, there are the edge currents which demonstrate the IQHE or the  
quantization of the Hall-conductivity in ideal cases \cite{kk}. In  
other words, the IQHE-behaviour is described by a $1+1$-D quantum  
theory.
On the other hand the mentioned Q-1-D superconductors should be  
described in our approach also as $1$-dimensional quantum systems by  
$1+1$-D quantum theories. Thus, in view of the fact that these  
superconductors demonstrate the mentioned quantum $1+1$-behaviour  
which is equivalent to the edge current-behaviour, they might  
demonstrate also the IQHE-behaviour. Moreover, the example of Q-1-D  
superconductors shows that not only the FQHE \cite{LafSC} but also  
the IQHE is related with the superconductivity.

 This considerations should be related with considerations in Ref.   
\cite{NNN} where also the main theoretical object is a $U(1)$ gauge  
potential which is introduced and gauged away frequently. It should  
be itresting to note here that, as in the case of Ref. \cite{NNN},  
every decomposition of an electromagnetic gauge potential is  
equivalent to its mathematical definition according to its gauge  
transformation property $A^{\prime} = A + \partial \phi$ which  
contains the pure gauge potential $\partial \phi$.

\medskip
On the other hand, the mentioned local equivalence of  
Ginzburg-Landau equations $J_{(em)} = \Delta A$ with the definition  
of pure gauge potential discussed above should be understood so that  
obviously $\Delta A = 0$ for pure gauge potentials. Furthermore, if  
we compare the strength of the increasing magnetic fields from weak  
and strong superconductivity to QHE which are described by  
London's, Ginzburg-Landau's and the Ohm's equations of  QHE  
respectively, then we find that: Depending on the preparation of the  
particle system, with increasing of the extrior magnetic field the  
$A$, $\Delta A$ an $dA$ of the pure gauge potential tends away from  
zero respectively but remain close to zero within the quantum  
mechanical uncertainty \cite{gauge} \cite{noph}. The strength of  
exterior magnetic field here is a quantum mechanical measure of the  
non-vanishing of $A$, $\Delta A$ and $dA = B$, if one recall that  
the energy uncertainty is given by the $\delta E = {\displaystyle  
{\frac{e\hbar B}{2m}}}$ \cite{unb}. This circumstance should be also  
understood according to the geometry on manifolds with different  
boundary structures \cite{kurz}.
Thus, depending on the preparation of the $2+1$-D particle system  
\cite{all 1} \cite{all 2} with increasing of the exterior magnetic  
field one has the empirical situations for which the London's-, the  
Ginzburg-Landau's- and the Ohm's equations of QHE are responsible  
respectively.
Furthermore, recall that the generality of the flux quantization  
concept for all of these three cases and also the generality of the  
BCS-theory for the first two cases result from the generality of the  
pure gauge electromagnetic potential which is according to our  
discussion the fundamental concept of all of them.

\medskip
In conclusion let us mention that for arbitrary locally constant  
$\sigma_H$ the variation of the action functional (3) with respect  
to $A_{\alpha}$ results in the following equations which are the  
generalization of the equations (4) or (5)-(6) for arbitrary  
$\sigma_H$:

\begin{equation}
J_{\alpha}^{(em)} = \epsilon^{\alpha\beta\gamma}\sigma_H  
\partial_{\beta} A_{\gamma}\;\;\;,
\end{equation}
\label{qhe}

where we used $J_{\alpha}^{(em)} =  
\lambda\epsilon_{\alpha\beta\gamma} \partial^{\beta}  
j^{\gamma}_{(e)}$ as given above according to the differential  
geometrical considerations about the current density involved in  
IQHE.

Therefore, if the $\sigma_H$ becomes quantized in view of a proper  
preparation of the particle system for larger exterior magnetic  
fields \cite{mein} \cite{kk}, then the spatial part of these  
equations become the Ohm's equations for the IQHE. Accordingly, if  
we perform a surface integral of the time component of (7) which is  
the constraint relation for constant mafnetic field $B$, we obtain  
the well known defining relation for the quantum Hall-conductivity  
in the quatum limit $\omega_c \tau \gg 1$ \cite{all 2}: $\sigma_H =  
{\displaystyle \frac{ne}{B}}$ which results in case of samples with  
proper prepared relations between $B$ and $n$ in quantized $\sigma_H  
= \nu{\displaystyle \frac{e^2}{h}}$.

This same results on IQHE including the Ohm's equations as  
equations of motion are obtained also from our  
Chern-Simons-Schroedinger-action for IQHE [12].

\bigskip
Footnotes and references


\begin{thebibliography}{100}
\bibitem{gauge}
In other words, one can transform the potential locally into the  
one equal to the gradient of the Phase of $\psi$. The global  
description of pure gauge character is given by the line integral of  
the potential which is equal to the mentioned phase. Furthermore,  
with pure gauge potential we mean everywhere a U(1)-type in multiply  
connected regions.
\bibitem {unb}
The uncertainty of $B$ should be considered as aresult of the  
energy uncertainty in Landau-levels i. e. $E = (N +  
\frac{1}{2}){\displaystyle {\frac{e\hbar B}{M_e}}}$.
\bibitem{noph}
See herefor [1d]. Furthermore, our model should be considered also  
in this category. Indeed the main consept in our $2+1$-D model of  
superconductivity is the compansated or in other words the absent  
Coulomb potential which is represented in our model by the pure  
gauge potential within the quantum mechanical uncertainity. It is  
related with the so called arbitrary weak attraction after  
compensation of the Coulomb repulsion in the phononic approach which  
is responcible quantum mechanically for the Cooper paires. It is  
also related with the energy uncertainty, the ground state energy  
and tunnel effects (see also the Josephson tunneling).
\bibitem{neu}
The concept of the $2$-dimensional polarized phase space of the  
$2+1$-D models is involved through the concepts of the Fermi-surface  
in the BCS-theory and in some other approaches in solid state  
physics. Recall that in view of the independency of the quantization  
from the polarization of the phase space also the Fermi-surface in  
momentum space can be the polarized phase space of the $2+1$-D  
system.
Furthermore, note also that for a Chern-Simons-action which is the  
typical action in $2+1$- dimensions the momentum and position space  
are the same.
\bibitem{text}
Recall further that in case of low dimensional superconductors the  
usual gap structure does not exist (see herefor Ref.[1d]).
\bibitem{kurz}
The detailed model will be discussed seperately
\bibitem{rot}
Recall from electrodynamics that a magnetic field is equivalent to  
a rotating coordinate system (see also the Ref. [1d]).
\end{thebibliography}
\end{document}